\documentclass[acmtog]{acmart}
\acmSubmissionID{261}

\usepackage{booktabs} 
\usepackage{amsmath}
\usepackage{amsbsy}

\acmPrice{15.00}

\setcitestyle{square}

\setcopyright{acmcopyright}\acmJournal{TOG}
\acmYear{2020}\acmVolume{39}\acmNumber{6}\acmArticle{179}\acmMonth{12} \acmDOI{10.1145/3414685.3417819}

\usepackage{microtype} 
\usepackage{xspace}
\usepackage{enumitem}

\usepackage{soul}
\usepackage{breakcites}
\usepackage{ellipsis} 
\usepackage[mathletters]{ucs}
\usepackage[utf8x]{inputenc}
\usepackage{mathtools}
\usepackage{amsbsy}
\usepackage{upgreek}
\usepackage{dblfloatfix}

\newcommand{\ra}[1]{\renewcommand{\arraystretch}{#1}}

\definecolor{lightbluishgrey}{rgb}{0.76078,0.88235,0.92157}

\usepackage{layouts}

\usepackage{wrapfig}

\newcommand{\refequ}[1] {Equation~(\ref{equ:#1})}

\newcommand{\reffig}[1] {Fig.~\ref{fig:#1}}

\makeatletter
\def\reffig{\@ifnextchar[{\@myreffigloc}{\@myreffignoloc}}
\def\@myreffigloc[#1]#2{Fig.~\ref{fig:#2}, \emph{#1}}
\def\@myreffignoloc#1{Fig.~\ref{fig:#1}}
\makeatother

\newcommand{\reftab}[1] {Table~\ref{tab:#1}}

\newcommand{\refsec}[1] {Section~\ref{sec:#1}}

\newcommand{\refalg}[1] {Algorithm~\ref{alg:#1}}

\let\mat = \mathbf
\usepackage{dsfont}

\newcommand{\R}{\mathbb{R}}
\newcommand{\Q}{\mathbf{Q}}
\newcommand{\vc}[1]{\mathbf{#1}}
\renewcommand{\C}{\mat{C}}

\renewcommand{\d}{\mathrm{d}}

\renewcommand{\b}{\vc{b}}
\newcommand{\cc}{\vc{c}}

\newcommand{\f}{\vc{f}}
\newcommand{\g}{\vc{g}}

\newcommand{\p}{\vc{p}}
\newcommand{\q}{\vc{q}}

\renewcommand{\u}{\vc{u}}
\newcommand{\vvv}{\vc{v}}

\newcommand{\x}{\vc{x}}

\newcommand{\A}{\mat{A}}
\newcommand{\B}{\mat{B}}
\renewcommand{\C}{\mat{C}}
\newcommand{\D}{\mat{D}}

\newcommand{\I}{\mat{I}}
\newcommand{\J}{\mat{J}}
\newcommand{\K}{\mat{K}}

\newcommand{\M}{\mat{M}}

\newcommand{\T}{\mat{T}}

\usepackage{siunitx}
\usepackage{graphicx}

\usepackage{wrapfig}

\makeatletter
\let\save@mathaccent\mathaccent
\newcommand*\if@single[3]{%
  \setbox0\hbox{${\mathaccent"0362{#1}}^H$}%
  \setbox2\hbox{${\mathaccent"0362{\kern0pt#1}}^H$}%
  \ifdim\ht0=\ht2 #3\else #2\fi
  }
\newcommand*\rel@kern[1]{\kern#1\dimexpr\macc@kerna}
\newcommand*\widebar[1]{\@ifnextchar^{{\wide@bar{#1}{0}}}{\wide@bar{#1}{1}}}
\newcommand*\wide@bar[2]{\if@single{#1}{\wide@bar@{#1}{#2}{1}}{\wide@bar@{#1}{#2}{2}}}
\newcommand*\wide@bar@[3]{%
  \begingroup
  \def\mathaccent##1##2{%
    \let\mathaccent\save@mathaccent
    \if#32 \let\macc@nucleus\first@char \fi
    \setbox\z@\hbox{$\macc@style{\macc@nucleus}_{}$}%
    \setbox\tw@\hbox{$\macc@style{\macc@nucleus}{}_{}$}%
    \dimen@\wd\tw@
    \advance\dimen@-\wd\z@
    \divide\dimen@ 3
    \@tempdima\wd\tw@
    \advance\@tempdima-\scriptspace
    \divide\@tempdima 10
    \advance\dimen@-\@tempdima
    \ifdim\dimen@>\z@ \dimen@0pt\fi
    \rel@kern{0.6}\kern-\dimen@
    \if#31
      \overline{\rel@kern{-0.6}\kern\dimen@\macc@nucleus\rel@kern{0.4}\kern\dimen@}%
      \advance\dimen@0.4\dimexpr\macc@kerna
      \let\final@kern#2%
      \ifdim\dimen@<\z@ \let\final@kern1\fi
      \if\final@kern1 \kern-\dimen@\fi
    \else
      \overline{\rel@kern{-0.6}\kern\dimen@#1}%
    \fi
  }%
  \macc@depth\@ne
  \let\math@bgroup\@empty \let\math@egroup\macc@set@skewchar
  \mathsurround\z@ \frozen@everymath{\mathgroup\macc@group\relax}%
  \macc@set@skewchar\relax
  \let\mathaccentV\macc@nested@a
  \if#31
    \macc@nested@a\relax111{#1}%
  \else
    \def\gobble@till@marker##1\endmarker{}%
    \futurelet\first@char\gobble@till@marker#1\endmarker
    \ifcat\noexpand\first@char A\else
      \def\first@char{}%
    \fi
    \macc@nested@a\relax111{\first@char}%
  \fi
  \endgroup
}
\makeatother

\usepackage[ruled,vlined]{algorithm2e}
\usepackage{etoolbox}
\usepackage{array}

\newcommand{\figs}{}
\def\figs/{figs/}

\DeclareMathOperator*{\argmin}{argmin}

\newcommand*\Bell{\ensuremath{\boldsymbol\ell}}

\usepackage[linewidth=1pt]{mdframed}

\newcounter{question}
\newcommand{\Question}[1]%
{%
  \stepcounter{question}%
  \textbf{\textcolor[rgb]{0.9,0.5,0.3}{\textsc{Question \thequestion}:}} %
  \emph{#1}%
}
\newcommand{\Hypothesis}[1]%
{%
  \textbf{\textcolor[rgb]{0.3,0.3,0.7}{\textsc{Hypothesis \thequestion}:}} {#1}
}
\newcommand{\Answer}[1]%
{%
  \textbf{\textcolor[rgb]{0.3,0.7,0.3}{\textsc{Answer \thequestion}:}} {#1}
}

\let\[\equation
\let\]\endequation

\newcommand{\udd}{\ddot{\u}}
\newcommand{\ud}{\dot{\u}}
\usepackage{bbm}

\renewcommand{\vec}[1]{\text{vec}(#1)}

\newcommand{\paragraphnodot}[1]{%
  \vspace*{1em}%
  \indent%
  \emph{#1}%
}

\newcommand*{\img}[1]{%
    \raisebox{-0\baselineskip}{%
        \includegraphics[
        keepaspectratio,
        ]{#1}%
    }%
}

\newcommand{\rig}{\mathbf{p}}
\usepackage{lipsum}

\begin{document}

\title{Complementary Dynamics}

\author{
  Jiayi Eris Zhang, Seungbae Bang, David I.W. Levin, Alec Jacobson
}
\affiliation{%
  \institution{University of Toronto}}

\begin{abstract}
We present a novel approach to enrich arbitrary rig animations with elastodynamic secondary effects.
Unlike previous methods which pit rig displacements and physical forces as
  adversaries against each other, we advocate that physics should complement
  artists' intentions.
We propose optimizing for elastodynamic displacements in the subspace orthogonal to displacements that can be created by the rig.
This ensures that the additional dynamic motions do not \emph{undo} the rig animation.
The complementary space is high-dimensional, algebraically constructed without manual oversight, and capable of rich high-frequency dynamics.
Unlike prior tracking methods, we do not require extra painted weights, segmentation into fixed and free regions or tracking clusters.
Our method is agnostic to the physical model and plugs into non-linear FEM
  simulations, geometric as-rigid-as-possible energies, or mass-spring models.
Our method does not require a particular type of rig and adds secondary
  effects to skeletal animations, cage-based deformations, wire deformers, motion
  capture data, and rigid-body simulations.
\end{abstract}

\ccsdesc[500]{Physical simulation}

\keywords{physically based animation, constrained simulation, character rigging, secondary motion, orthogonality}

\begin{teaserfigure}
  \centering
  \includegraphics[width=\linewidth]{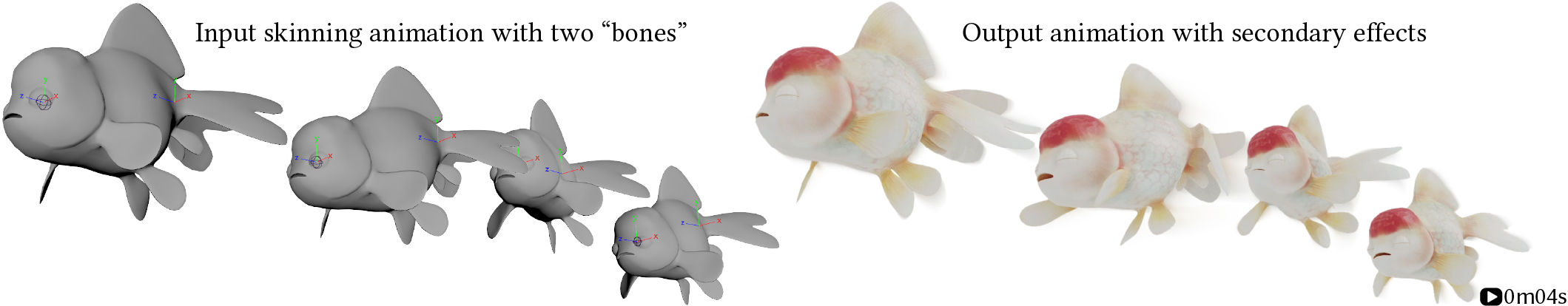}
  \caption{
    \label{fig:teaser}
    Animation artists use low-dimensional rigs to control the primary motion of
    their characters.
    Our displacement filtering optimization adds elastodynamic secondary
    effects \emph{orthogonal} to the rig's subspace, ensuring that interesting
    effects emerge (right) but do not compete with or undo the artist's original
    animation (left). Throughout the figures in our paper, the
    \protect\img{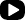} symbol indicates a corresponding clip in the supplemental
    video.
  }
\end{teaserfigure}

\maketitle


\section{Introduction}
The performance of elastodynamics simulation algorithms for character animation
has improved by leaps and bounds in the past few years. 
%
%
More than ever, the major bottleneck in realistic animation creation is artistic
control. 

Controlling physical effects by meticulously setting up initial conditions and
material parameters alone is a non-starter.
Ideally, an artist could use friendly interfaces to animate dominant effects
(e.g., keyframing, mocap, skinning) and rely on physical simulation to add
secondary effects.
Space-time constraint systems use physics to interpolate sparse key poses, but
this can be awkward, inflexible, and computationally intractable.
Existing tracking methods can also be frustrating: either physics does not have
enough freedom to add interesting effects or physics has too much freedom and
\emph{undoes} the artist's intentions.

A paradox seems to appear.
The artist's rig displacements cannot be treated as hard constraints: otherwise
physics has no room for secondary effects.
Meanwhile, physics can not have too much freedom, so as to undo the artist's
work.

We observe that these goals --- creative primary effects and physical secondary
effects --- are not contradictory, but rather they are \emph{complementary}.
In this paper, we show that this is true not just philosophically, but
also \emph{algebraically}.
We propose a displacement filtering approach that enables physically based
secondary dynamics lying solely in the \emph{orthogonal complement} of the
artist's animation.
Our physical simulation can add wiggles and jiggles that respond to external
forces by superimposing displacements that the artist \emph{strictly} could not
have made herself with the given rig.
That is, the physical simulation cannot add a displacement that could have
otherwise been added by the artist.
This implements a \emph{contract} with the artist ensuring that their creative
intentions are not undone by the physical dynamics, while ensuring that the
simulation has enough degrees of freedom to add interesting effects.

In contrast to previous tracking or constraint-based methods, our method does
not require a careful segmentation, remeshing, or masking of the input geometry.
%
%
Our method is plug-and-play with existing elastodynamics methods, and as a
result inherits their real material parameters and non-linear effects.
Our method is agnostic to the elastic model, applying both to continuum models
and simple mass-spring systems.
We require only first-order differentiability of the input animation with respect to the
artist's parameters allowing us to add secondary effects to animations created
with linear blend skinning, motion capture, cage-based animation
and non-linear rigs such as dual quaternion skinning and wire deformers.
We demonstrate prototypical use case for 2D cartoon and 3D character animation.
Secondary effects can be triggered by contacts and collisions or other
environmental forces.
As an exciting special case, our method can enrich rigid-body simulations, where
we interpret each object as a keyframed motion.
Even in the absence of environmental forces, we show how to effortlessly allow
the rig to inject momentum into the secondary effects.
This affords extremely low setup virtual puppetry (see
\reffig{teaser}). 

\section{Related Work}
\label{sec:related}
The past few years have culminated in major robustness and performance
improvements for elastodynamic simulation (see, for example,
\cite{SmithGK18,KimGI19,BarbicZ11,TengMDK15,WangY16,liu2017quasi,WangWFTW18,PengDZGQL18,DinevLLTK18,XianTL19,ChenBWD018,ChenBOBHD19}).
We benefit from this tremendous body of work, and our contributions could be
combined with each new improvement.
We do not present a new model of elasticity nor a performance optimization of an
existing one.
Rather, our goal is to determine the right mathematical interface between
creative artistic animations and physical secondary effects.
So, while exploring all combinations with recent advances in
elasticity simulation is interesting, it is outside the scope of the work
presented here and we leave the problem of ``real-time
complementary dynamics'' to future work.

Using physical simulation for secondary effects in computer graphics is a
classic problem with a broad literature.
We can categorize previous works in relation to ours by examining the
``contract'' they make with the user.
The most extreme contracts being: 
no physical effects, leaving full control \emph{and responsibility} to the
artist to create a realistic animation (e.g., skinning, blendshapes or
keyframing); and full physical simulation, with almost no direct artistic
control besides setting up initial conditions and material parameters.
There have been many interesting prior works spanning the spectrum between these
extremes.

\subsection{Positional Constraints}
Skeletal skinning is a popular method for character animation
\cite{skinningcourse}, but lacks procedural secondary effects.
Instead of mapping bone transformations to the skin via static weights, the
bones of the skeleton can be interpreted geometrically, as embedded rigid solids
within the elastic solid of the character's interior
\cite{CapellGCDP02,KimJ11,MullerDMJC02,LarbouletteCA05,ShiZTDBG08,McAdamsZSETTS11,KavanS12,KomaritzanB18,KomaritzanB19,LiLK19}.
The skeleton's animation is applied as temporally varying Dirichlet boundary
conditions to the dynamic (e.g., FEM) simulation.
The contract with the user is thus reduced to the geometric animation of the
piecewise-rigid embedded skeleton.
This leaves the control of the ``skin'' (i.e., the visual exterior) delegated to
the physical simulation.
Achieving a desired silhouette \cite{illusionoflife} or controlling soft surface
features such as facial expressions can be difficult or impossible with rigid
internal bones alone.

\begin{figure}
\includegraphics[width=\linewidth]{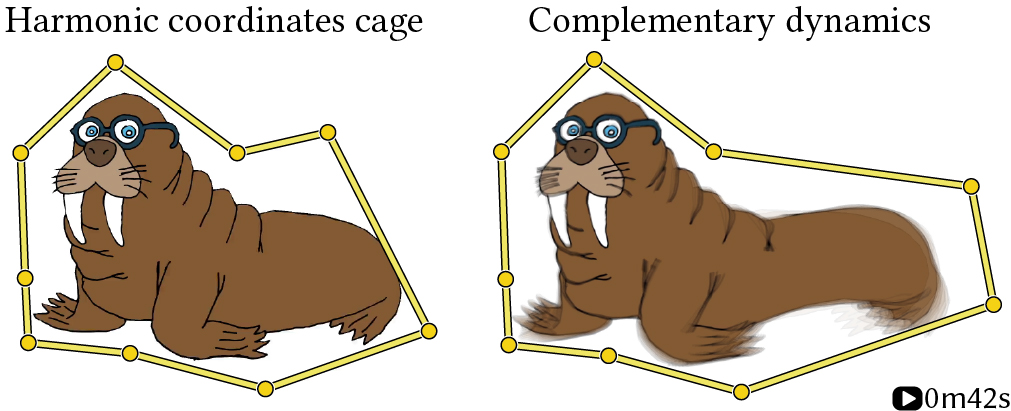}
\caption{\label{fig:walrus-cage} 
A cartoon walrus jiggles at frequencies that the cage (Harmonic Coordinates)
  cannot create on its own.
  }
  \vspace*{-0.4cm}
\end{figure}

\begin{figure}
\includegraphics[width=\linewidth]{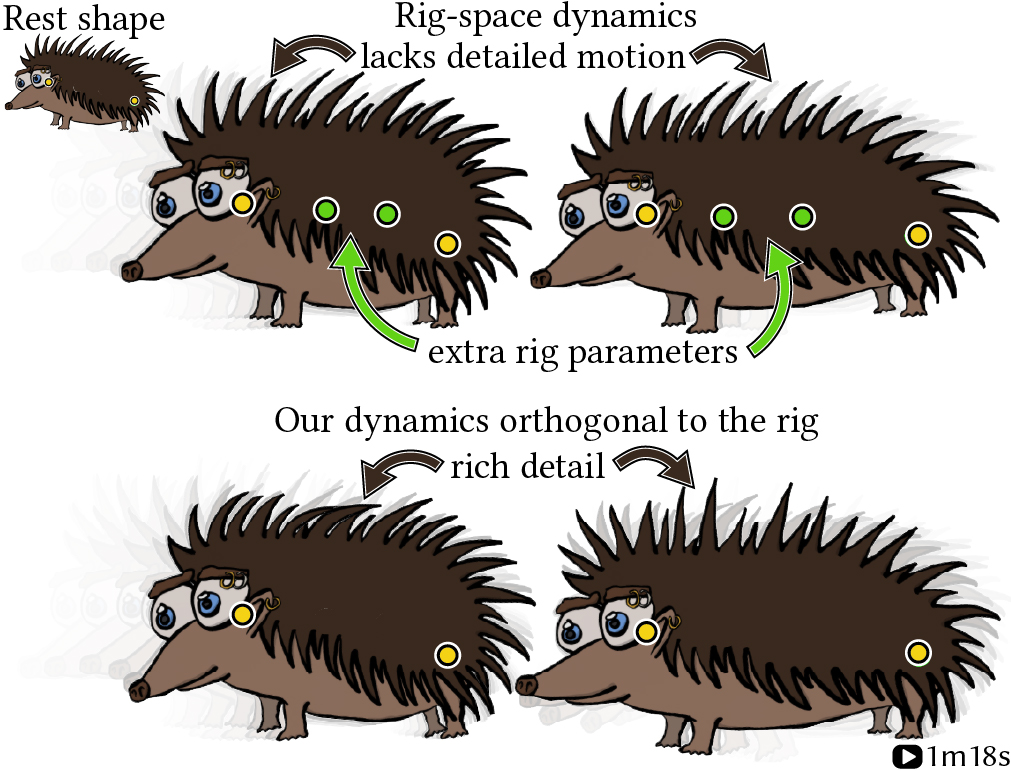}
\caption{\label{fig:hedgehog} 
  Simulation in the subspace of the rig (e.g., \cite{hahn2012rig}) is limited by
  the expressivity of the rig, which is typically designed for primary motions.
  If the artist plans to control the entire rig, then extra parameters need
  to be added for rig-space physics to have an effect.
  Our secondary effects lie in the space \emph{orthogonal} to the rig
  and require no extra rigging.
  }
  \vspace*{1em}
\includegraphics[width=\linewidth]{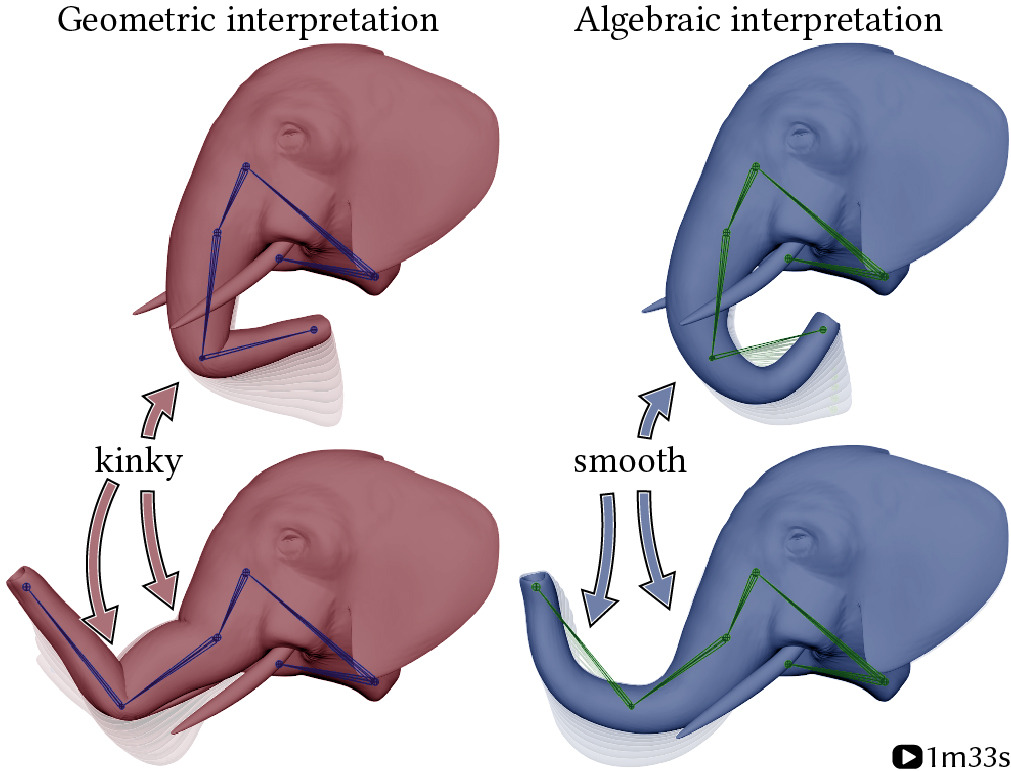}
\caption{\label{fig:elephant-head} 
Previous methods interpret input skeletal rigs as ``trees of rigid bars''
  embedded in the elastic shape (e.g., \cite{CapellGCDP02,LiLK19}).
  But often the skeleton is just a hierarchy metaphor for controlling a
  smooth subspace.
  Our interpretation of the rig is through its action on the shape, not its
  geometric handle locations.
  }
  \vspace*{-0.3cm}
\end{figure}

This literal interpretation of the skeletal control rig requires meaningful
``geometric bones'' for each rig parameter, thus does not directly apply to:
cage-based deformation (\reffig{walrus-cage}), blendshapes, abstract rigs, or
rigid-body controllers.
Many \emph{hierarchical} rigs use the skeleton metaphor, but are not intended to
be geometrically interpreted as ``trees of rigid bars'' inside the shape.
In \reffig{elephant-head}, two ``bones'' with smooth skinning weights control
the soft trunk of an elephant. 
Interpreting the rig as a geometric skeleton so as to impose fixed value
constraints results in awkward kinks.
In contrast, our approach interprets the rig \emph{algebraically} and
augments the original smooth skinning deformation with smooth secondary
dynamics.

In lieu of geometric bones, the user could be asked to specify which regions
of an input animation should be \emph{fixed} and which should be \emph{free}
\cite{kim2017data,li2016enriching,KozlovBBTBG17}.
In many cases, this is a non-trivial segmentation task, perhaps more difficult
than modeling the input's surface geometry.
Fixing too much of the interior leads to diminished secondary effects; fixing
too little or too deep will not allow the artistic intention to diffuse to the
surface.
\citet{li2013thin} treat the entire geometry of the input animation as fixed
except for secondary skin sliding effects tangential to the surface, however
this requires a high quality UV parameterization with seam bookkeeping.
\citet{MalgotMH2011} use a kinetic filter at the velocity level to add local,
physics-driven motion to underlying coarse discretizations. In contrast to our
method which allows global interactions, their method only allows adding
localized motion, making it unsuitable for many tasks in animation. 
It also requires additional positional constraints to avoid constraint-drift
caused by linearizing constraints at the velocity level.

\begin{figure}
  \includegraphics[width=\linewidth]{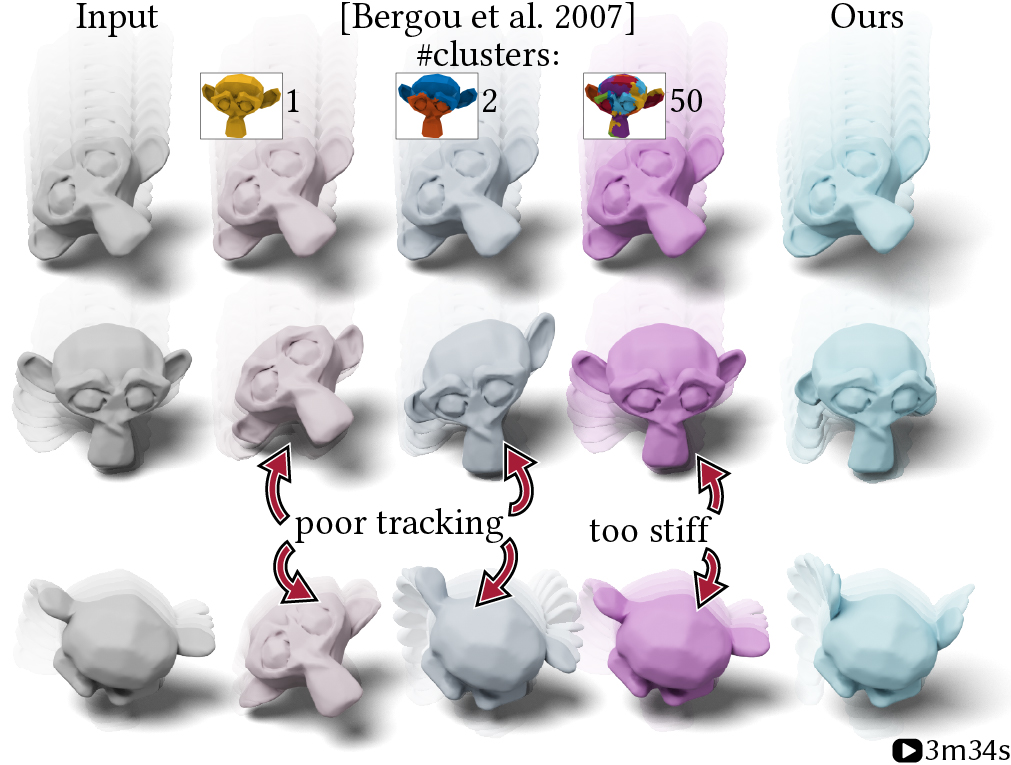}
  \caption{\label{fig:suzanne-tracks}
  \citet{bergou2007tracks} track input animations by weak constraints defined
  over patches. The output is sensitive to the distribution and number of patches.
  Problems occur at either extreme.
  Finding a good balance is an additional burden for the user. 
  Our method does not require
  clustering.}
\end{figure}

Instead of a geometric rig or \emph{fixed} inner core, \citet{bergou2007tracks}
input an existing coarse mesh animation and then require a high-resolution
simulation match its positions in a weak sense (integrated over pre-defined
patches).
Again, the user must find a good segmentation.
\reffig{suzanne-tracks} shows that too few patches allow the tracked simulation
to deviate from the input and too many prohibit secondary effects.


\subsection{Spacetime Constraints}
Keyframing is another popular animation metaphor, traditionally lacking
secondary effects.
Instead of interpolating rig parameters between key poses, the key poses can be
interpreted as sparse constraints on a spacetime optimization to find the most
physically plausible animation \cite{WitkinK88}.
This changes the simulation from an instantaneous integration problem into a
coupled problem over all timesteps; many of the recent advances have sought to
improve the runtime performance of this optimization
\cite{BarbicSP09,hildebrandt2012interactive,schulz2014animating,LiHG0BD14}.
Spacetime constraints make it very easy to strike a specific pose.
On the other hand, an artist's desired arc of motion may not be 
physically efficient and therefore avoided by the optimization.
The artist may have to specify more and more poses to cajole the physical
interpolation onto a desired motion.
It is hard to tell in advance how many poses are needed.
This contract limits the creativity of
animators to providing sparse poses giving physics full power over \emph{motion},
even primary motions, not just secondary effects.

%
%

\subsection{Rig-Space Physics}
\label{sec:rigspace}
%
\emph{Rig space physics} constrains the displacements of secondary effects to
lie in the subspace spanned by the artists' rig 
\cite{hahn2012rig}.
This has the immediate performance advantages of a reduced deformable model
\cite{BarbicJ05,DerSP06,GillesBFP11,XuB14} and has proven useful as a form of
regularization during performance capture \cite{Ribera2017,LiuLDT20}.
For character animation, the contract with the artist is explicit: the artist
segments the rig parameters into free and fixed.
The failure modes are two sides of the same coin.
Since physics can only make motions spanned by the rig, interesting secondary
effects may require augmenting the rig with new auxiliary degrees of freedom
\cite{JacobsonBKPS12,Wang2015}.
Alternatively, the artist loses control over parameters delegated to physics,
which may prevent realization of the artist's intent.
When acting in the same subspace, the artist and physical simulation are in
adversarial roles.
Instead, they could be working in harmony to control primary and secondary
effects respectively in complementary subspaces.

\begin{wrapfigure}[8]{r}{1.68in}
  \includegraphics[width=\linewidth,trim={6mm 0mm 0mm 4mm}]{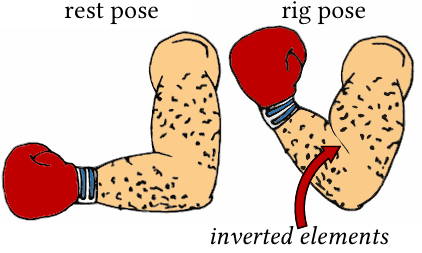}
  \label{fig:boxer-inverted}
\end{wrapfigure} 
\subsection{Modifying the Reference Configuration}
\begin{figure}
  \includegraphics[width=\linewidth]{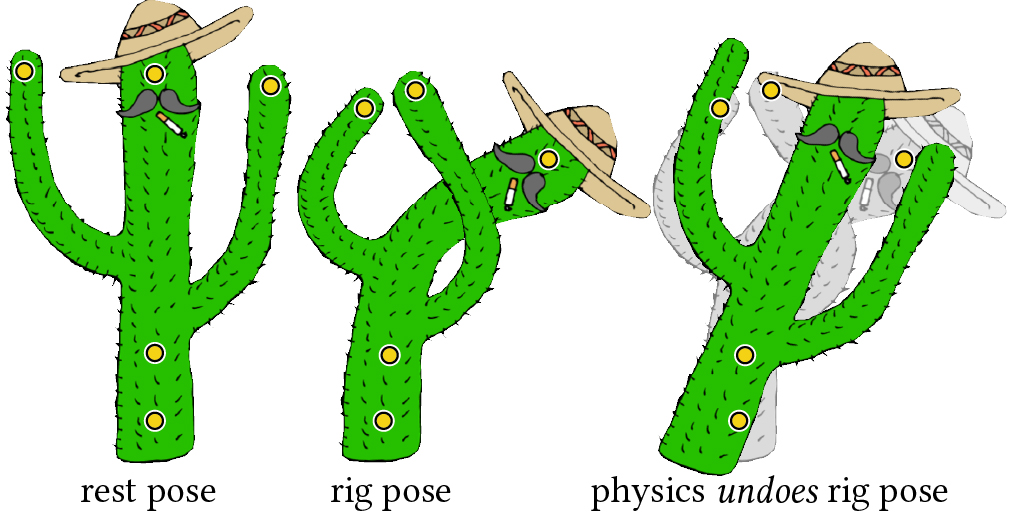}
  \caption{\label{fig:cactus-physics-undoes-rig}
  If secondary effects minimize an elastic energy, but no
  constraints are imposed, then the physics will instantly \emph{undo} the
  desired rig pose.
  }
\end{figure}

A core common problem is that giving too much freedom to the physics simulation
will \emph{undo} the artists' intended primary animation (see
\reffig{cactus-physics-undoes-rig}).
Minimizing an elastic energy exerts forces pushing the shape back to its
reference
configuration \emph{away} from the artist's pose.
Treating the current rig pose at any moment in the animation as the reference
configuration will prevent this
\cite{AngelidisS07,Ma2011,hahn2012rig,xu2016pose,KozlovBBTBG17}; almost as if
the rig is a muscle \cite{CorosMTSSG12}.
This approach is agnostic to how the pose is created (skinning,
blendshapes, etc.).
Unfortunately, changing the reference configuration can cause catastrophic
failure if the rig pose creates a physically impossible or infinite-energy
configuration (see inset).
Moreover, in 3D the input rig typically only controls the surface, so the rig
pose must be extrapolated to the interior to redefine a valid rest pose
\cite{hahn2012rig} --- this fragile process must be conducted every
frame of the animation.

\section{Method}
%
\label{sec:method}
Given a rigged 2D or 3D shape, our method takes as input the set of rig
parameters for the next animation frame and outputs a set of displacements
adding dynamic secondary effects.
For now, we assume the input shape $Ω∈\R^d$ is represented by a mesh with $n$
vertices.
We treat the \emph{rig} as a function $\u^r$ mapping a small set of $m$
parameters gathered in vector $\rig ∈ \R^m$ to \emph{displacements} of the mesh
vertices: $\u^r : \R^m → \R^{dn}$. At some time $t$ during the animation, we
write the current rig parameters as $\rig_t$ and the current rig displacements
as $\u^r(\rig_t)$ or simply $\u^r_t$.
Our goal is find \emph{complementary} displacements $\u_t^c$ to add
dynamic secondary effects to the rigged input. The final displacements are
the sum of the rig and complementary displacements:
\begin{equation}
  \label{equ:sum}
  \u_t = \u_t^r + \u_t^c.
\end{equation}

We define two 
guidelines
for finding appropriate $\u^c$ values:
%
\begin{enumerate}[label=(\Roman*)]
  \item $\u^c$ should react to internal and external forces, and 
  \item $\u^c$ should not \emph{undo} the rig displacements $\u^r$.
\end{enumerate}
Guideline (II) is trivially satisfied by $\u^c=0$, but then of course we get no
reaction to physics. Similarly, in the absence of inertial or external forces,
guideline (I) is satisfied by setting $\u^c
= -\u^r$ returning the shape to its rest pose, but this (completely) undoes the
rig displacements.
Let us first describe our general physical model, then how we model
rig \emph{complementarity}, and finally consider specific instances for common
elastic potentials and rig functions.

\subsection{Dynamic Simulation}
Following many previous works (e.g., \cite{MartinTGG11,liu2013fast,GastSSJT15}),
we employ an implicit Euler integration of Newton's Second Law of Motion in terms
of displacements. For a given moment in time, $t$, the current displacements
$\u_t$ are the result of a possibly non-linear optimization problem:

\begin{align}
  \label{equ:full}
  \u_t = \argmin_{\u_t} \ E_t(\u_t),
\end{align}
where $E_t$ changes over time due to momentum and external forces and is defined
as a sum of potentials:
\begin{align}
  \label{equ:Et}
  E_t(\u_t) =  
  \underbrace{Φ(\u_t)}_\text{potential energy} + \underbrace{\frac{h^2}{2}
  \udd_t^\top \M \udd_t}_\text{momentum term} + \underbrace{-\u_t^\top
  \f(\u_t),}_\text{external work}
\end{align}
where $h>0$ is the time step value, $\M ∈ \R^{dn×dn}$ is the mass matrix,
$\f:\R^{dn}→\R^{dn}$ defines external forces, and we
use dot notation to denote temporal finite differencing:
\begin{gather}
  \udd_t := \frac{\ud_t - \ud_{t-h}}{h} \quad \text{ and } \quad
  \ud_t := \frac{\u_t - \u_{t-h}}{h}.
\end{gather}

Substituting \refequ{sum} into \refequ{full} results in a generic optimization
problem over the complementary displacements $\u^c$:
\begin{align}
  \label{equ:fulluc}
  \u^c_t = \argmin_{\u^c_t} \quad E_t(\u^r_t + \u^c_t).
\end{align}
Unconstrained, this optimization will \emph{undo} the rig displacements $\u^r_t$
(see \reffig{cactus-physics-undoes-rig}).
Requiring $\u$ to \emph{track} $\u^r$ in a least-squares sense (e.g., adding a
$\|\u-\u^r\|²$ or equivalently $\|\u^c\|²$ term) will prevent undoing the rig
displacements but also damp out high-frequency motions. 
These are motions that the rig cannot create, so they would otherwise be welcome
secondary effects (see \reffig{suzanne-tracks}).
Instead, we now introduce constraints to prevent \emph{exactly} the dynamics
lying in the space spanned by the rig.

\subsection{Rig Orthogonality Constraints}
Our goal is to constrain the complementary displacements $\u^c$ to only those
displacements that \emph{could not be created by the rig} $\u^r$.
That is, physics should not presume to \emph{take over} any controls from the
artist using the rig.
Given current rig parameters $\rig_t$, we can verify whether any candidate
displacements $\u^c_t$ satisfies this criteria by ensuring that projecting to the
closest rig displacement simply rediscovers $\rig_t$:
\begingroup
\renewcommand{\q}{\rig}
\begin{equation}
  \label{equ:argmin}
  \argmin_\q \frac{1}{2} \left\| \u^r(\rig_t)+\u^c_t - \u^r(\q)\right\|_\M^2
  = \rig_t,
\end{equation}
where $\|\x\|_\M^2 = \x^\top \M \x$.
This should not be confused for a definition of $\rig_t$, rather it is a (currently
non-linear) equality constraint on $\u^c_t$.
This constraint appears daunting as it involves $\argmin$ and the possibly
non-linear rig function $\u^r$.

Let us consider the first-order necessary conditions of the $\argmin$
operation with respect to $\q$ evaluated at the right-hand side $\rig_t$:
\begin{align}
\left. \frac{∂ \frac{1}{2} \left\| \u^r(\rig_t)+\u^c_t - \u^r(\q)\right\|_\M^2
  }{∂ \q} \right|_{\rig_t} &= \mathbf{0}, \\
\left. \frac{∂ \left( \u^r(\rig_t)+\u^c_t - \u^r(\q)\right) }{∂ \q} \right|_{\rig_t}^\top \M \left( \u^r(\rig_t)+\u^c_t - \u^r(\rig_t)\right)
  &= \mathbf{0}, \\
  \label{equ:Jconst}
  \underbrace{\left. \frac{∂ \u^r }{∂ \q} \right|_{\rig_t}^\top}_{\J_t^\top} \M \u^c_t
  &= \mathbf{0}.
\end{align}
\endgroup
The final expression is \emph{linear} in the unknown complementary displacements $\u^c$.
The rig Jacobian matrix $\J_t ∈ \R^{dn × m}$ in general changes with each time
step, but does not depend on the unknown complementary displacements $\u^c$.

We assume that the input rig function $\u^r$ is differentiable.
This is reasonable as most rigs are intended to create smooth
animations.
This is also milder and less computationally intensive than
previous methods: \citet{hahn2012rig} require second derivatives.

\begin{wrapfigure}[10]{r}{1.5in}
  \includegraphics[width=\linewidth,trim={6mm 0mm 0mm 4mm}]{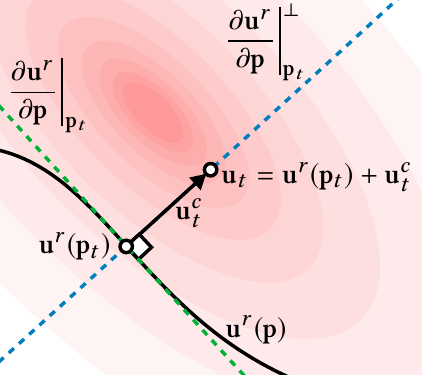}
  \label{fig:orthogonality-notation}
\end{wrapfigure} 
This constraint can read as requiring that $\u^c$ lies in the orthogonal
complement of the rig, to first-order.

\begin{figure}
  \includegraphics[width=\linewidth]{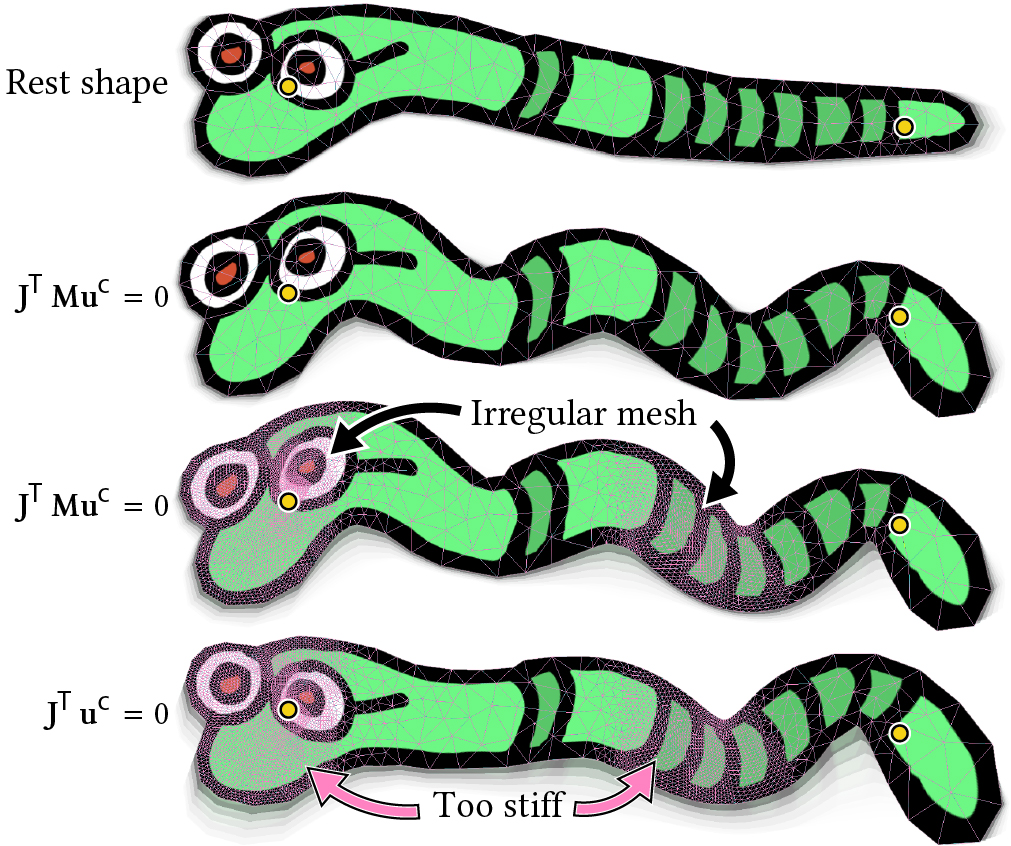}
  \caption{\label{fig:worm-modes} 
  Complementary vibrations jiggle through a cartoon worm (top) with a coarse
  mesh (top-middle). Including the mass matrix $\M$ in the constraint  ensures
  discretization independence (bottom-middle). Omitting it causes densely meshed
  regions to be overly stiff.
  }
\end{figure}

The mass matrix $\M$ appears in \refequ{Jconst} due to the integrated measure in
\refequ{argmin}.
Omitting this mass matrix results in secondary effects that are biased by the
mesh discretization (see \reffig{worm-modes}).

\subsection{Constrained Simulation}
Our general algorithm involves differentiating the rig at the current pose,
performing one constrained optimization integration step, and then summing the
displacement contributions (\refalg{sim}).
\begin{algorithm}[h!]
  \caption{Complementary Dynamics Simulation: given $\rig_t$}
  \label{alg:sim}
  $\J_t ← \left.\frac{d \u^r_t}{d \rig}\right|_{\rig_t}$ \\
  $\u_t^c ← \argmin\limits_{\u_t^c} E_t(\u^r_t+\u^c_t) \text{ subject to } \J_t^T
  \M \u^c_t = 0$ \\[0.3em]
  $\u_t ← \u^r_t+\u^c_t$
\end{algorithm} 

To demonstrate the effectiveness and generality of this method, let us consider
various choices of rig functions ($\u^r$) and various choices of elastic
potentials ($Φ$ within $E_t$ in \refequ{Et}).

\paragraph{\textbf{Linear Elasticity}}
\label{sec:linear-elasticity}
As a warm-up, let us start with linear elasticity which is captured by a
quadratic potential:
\begin{equation}
Φ_\text{linear}(\u) = \frac{1}{2} \u^\top \K \u,
\end{equation}
where $\K ∈ \R^{dn × dn}$ is the \emph{constant} stiffness matrix.

Using the Lagrange multiplier method, the optimal complementary displacements
$\u_t^c$ can be found by solving the linear system:
\begin{equation}
  \label{equ:lm}
  \hspace*{-0.1cm}
  \begin{bmatrix}
    \K + \frac{\M}{h²} & \M^\top \J_t \\
    \J_t^\top \M & \mathbf{0}
  \end{bmatrix}
  \begin{bmatrix}
    \u_t^c \\ \lambda
  \end{bmatrix}
  =
  \begin{bmatrix}
    -\K \u^r_t - \frac{\M}{h} \left( \frac{\u^r_t - \u_{t-h}}{h} -
    \ud_{t-h}\right) + \f_t
    \\ 
    \mathbf{0}
  \end{bmatrix}
\end{equation}
where the Lagrange multiplier values $λ∈\R^m$ may be discarded.

As a quadratic function, the Hessian of $Φ_\text{linear}$ is constant and
\refequ{lm} can be interpreted as a single (perfect) Newton–Raphson method
iteration where $\K = ∂²Φ/∂{\u^c}²$ and $\K (\u^r_t + \u^c_t) = ∂Φ/∂\u^c$.

\paragraph{\textbf{Non-Linear Elasticity}}
Linear elasticity is a poor model for large displacements.
A popular non-linear replacement is neo-Hookean elasticity, a non-linear model,
$Φ_\text{neo}$ (see, e.g., \cite{Sifakis12}), though the derivations below
generalize to many non-linear models.
We optimize the complementary displacements $\u^c_t$ again by using the Lagrange
multiplier method, but now applied repeatedly during each iteration of a
Newton-Raphson's algorithm. 
Each iteration the potential $E$ is approximated to second-order using a Taylor
expansion, requiring the first and second derivatives of the neo-Hookean
potential (\refalg{nm}).
\begin{algorithm}[h!]
  \caption{Modified Newton's Method: given $\u^r_t$, $\J_t$}
  \label{alg:nm}
  $\u^c_t ← \u^c_{t-h}$\\
  \Repeat{$s < ε$}{
    $\g,\K ← ∂Φ/∂\u^c,∂²Φ/∂{\u^c}²$ \\
    $\Q ← \K + h²\M $\\
    $\Bell ← -\g + \frac{\M}{h} \left( \frac{\u^r_t - \u_{t-h}}{h}
      - \ud_{t-h}\right) + \f_t $\\
    $\C ← \J_t^\top \M$\\
    Solve
    $
    \begin{bmatrix}
      \Q & \C^\top \\
      \C & 0
    \end{bmatrix}
    \begin{bmatrix}
      \x \\
      \lambda
    \end{bmatrix}
    =
    \begin{bmatrix}
      \Bell \\
      \mathbf{0}
    \end{bmatrix}
    $\\
    $s ← \text{line search from $\u_t^c$ toward $\x$ according to $E_t$}$\\
    $\u_t^c ← \u_t^c + s(\x-\u_t^c)$
  }
\end{algorithm}

\begin{wrapfigure}[11]{r}{1.0in}
  \includegraphics[width=\linewidth,trim={5mm 0mm 0mm 12mm}]{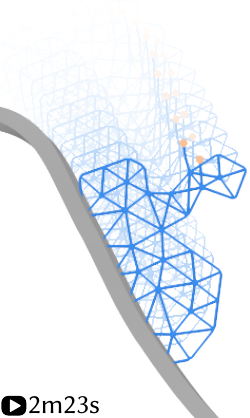}
  \label{fig:roller}
\end{wrapfigure} 
\paragraph{\textbf{Local-Global Energies}}
\label{sec:local-global}
While Newton's method can be applied to a large class of energies, a popular
style of optimization for elastic simulations are so-called local-global solvers
(e.g.,
\cite{SorkineA07,JacobsonBKPS12,liu2013fast,bouaziz2014projective,Wang2015,KovalskyGL16}).
Let us consider the mass-spring solver of \citet{liu2013fast}.
This method introduces auxiliary ``local'' variables representing the
fixed-length direction of each spring so that the elastic potential becomes
quadratic with respect to the ``global'' displacement variables.
The optimization alternates between updating the local variables via vector
normalization and solving a linear system to update the global variables.
We can immediately bootstrap this optimization by enforcing $\J_t^\top \M \u^c_t =
0$ as a linear equality constraint during the global step, leaving the local
step intact.
The inset figure shows mass-spring secondary-effects enriching a keyframe
animation of a roller coaster.

\subsection{Rig Derivatives}
\label{sec:rigd}
Our method requires differentiating the rig displacements with respect to the
rig pose to build the Jacobian matrix of partial derivatives, $\J = \frac{\d
\u^r}{\d \rig} ∈ \R^{dn×m}$.
Many popular rigging methods are \emph{linear} in their rig parameters, so their
Jacobian
matrix is \emph{constant}.

\paragraphnodot{\textbf{Linear Blend Skinning}} computes displacements at each vertex with
rest position $\vvv_i ∈ \R^d$ as a
weighted average of $k$ affine ``bone'' transformations ($\T_{j} ∈ \R^{d × (d+1)}$
for each bone $j$):
\begin{equation}
  \label{equ:lbs}
  \u^\text{lbs}_i = \left(\sum_{j=1}^k w_{ij} \T_j 
    \begin{bmatrix} \vvv_i \\ 1 \end{bmatrix}\right) - \vvv_i,
\end{equation}
where the scalar per-vertex-per-bone skinning weights $w_{ij}$ are constants at
pose time.

If we vectorize each matrix $\T_j$ into $\rig_j = \vec{\T_j^\top} ∈ \R^{d(d+1)}$, then
we can write linear blend skinning displacements in \refequ{lbs} as a sum of 
matrix multiplications:
\begin{equation}
  \u^\text{lbs}_i = \left(\sum_{j=1}^k 
  \underbrace{w_{ij} (\I_d \otimes [\vvv_i^\top\ 1])}_{\J_{ij}} \rig_j
  \right) - \vvv_i
\end{equation}
where $\I_d ∈ \R^{d×d}$ is the identity matrix and $\A \otimes \B$ indicates the
Kronecker product of $\A$ and $\B$.
Concatenating all of the $\rig_j$ vectors into one tall vector of rig parameters
$\rig∈\R^{kd(d+1)}$, we can write the linear blend skinning displacement function
as a matrix multiplication and vector subtraction:
\begin{equation}
  \label{equ:lbsJ}
  \u^r = \underbrace{\begin{bmatrix}
    \J_{11} &\cdots & \J_{1k} \\
    \vdots & \vdots & \vdots \\
    \J_{n1} & \cdots & \J_{nk}
  \end{bmatrix}}_{\J} \rig - \vvv.
\end{equation}
As such, the matrix $\J$ is immediately revealed to be the rig's constant
Jacobian. 

\paragraphnodot{\textbf{Affine Body}} A very useful special case of linear
blend skinning is to consider that the entire object is controlled by a single
affine transformation (and all weights are one $w_{i1} = 1$). This is a very
common control metaphor in keyframing software such as \textsc{Maya},
\textsc{Blender}, and \textsc{After Effects}. In this case the matrix expression
in \refequ{lbsJ} reduces to:
  \begingroup
  \renewcommand{\arraystretch}{1.1}
\begin{equation}
  \label{equ:affineJ}
  \u^\text{affine} = 
  \underbrace{\left(
    \I_d \otimes
  \begin{bmatrix}
    \vvv_1^\top & 1 \\
    \vvv_2^\top & 1 \\
    \multicolumn{2}{c}{\vdots} \\
    \vvv_n^\top & 1
  \end{bmatrix}\right)}_{\J} \p - \vvv
\end{equation}
\endgroup

\paragraphnodot{\textbf{Cage-based Deformation}} using generalized barycentric
coordinates (e.g., \cite{JoshiMDGS07}) is another popular special case of linear
blend skinning \cite{skinningcourse}.
Generalized barycentric coordinates represent the new position of each vertex as
a weighted average of the $k$ posed cage vertex positions $\cc_i ∈ \R^d$:
\begin{equation}
  \label{equ:cage}
  \u^\text{cage}_i = \left( \sum\limits_{j=1}^k w_{ij} \cc_i \right) - \vvv_i.
\end{equation}
Concatenating all cage vertex positions into $\rig ∈ \R^{kd}$, then $\u^\text{cage}
= \J \rig - \vvv$ where $\J$ simply contains the cage coordinates ($\J_{ij} =
w_{ij}$).

\paragraphnodot{\textbf{Blendshapes}} displace vertices by taking weighted
averages of $m$ sculpted poses of the entire model:
\begin{equation}
  \u^\text{blend} = \left( \sum_{j=1}^m w_j \b_j \right) - \vvv,
\end{equation}
where $\b_j ∈ \R^{dn}$ contains the vertex positions of the $j$th sculpted pose
and $w_j$ is the corresponding weight. In this case, the weights $w_j$ are the rig
parameters that vary over time. Written in matrix form, blendshape deformation
reveals its constant Jacobian:
\begin{equation}
  \u^\text{blend} = 
  \underbrace{
  \begin{bmatrix}
    \b_1 \cdots \b_m
  \end{bmatrix}
  }_{\J}  
  \underbrace{
  \begin{bmatrix}
    w_1 \\ \vdots \\ w_m
  \end{bmatrix}
  }_{\rig}  
  - \vvv.
\end{equation}

For any of the previous linear rig functions, the Jacobian matrix $\J$ can be
pre-computed once at the beginning of the animation as it does not depend on the
changing parameters $\rig_t$.
When used in combination with \emph{Linear Elasticity} or \emph{Local-Global
Energies}, the resulting large sparse system matrix also remains constant and
its $LDL^T$ factorization can be precomputed.

\begin{figure}
  \includegraphics[width=\linewidth]{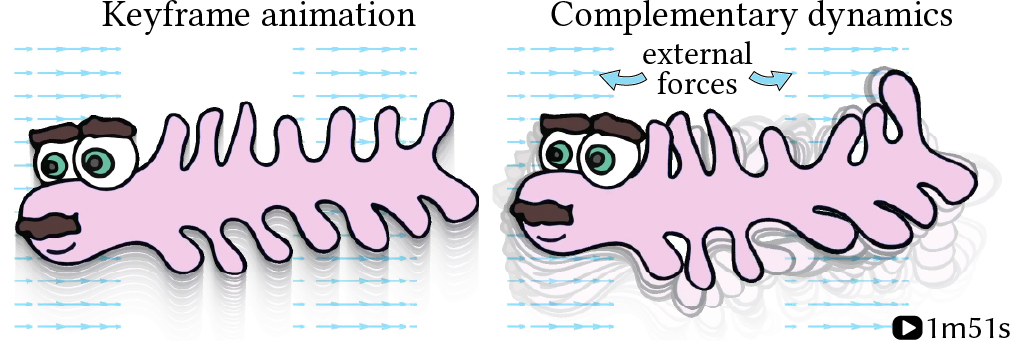}
  \caption{\label{fig:amoeba} 
  Keyframe animation provides the primary motion of this amoeba cartoon, but
  secondary responses to externals forces bring it to life.
  }
\end{figure}

\paragraph{\textbf{Non-Linear Rigs}}
\label{sec:non-linear-rigs}
Other popular rigging systems are \emph{non-linear} in their rig parameters $\rig$,
and therefore the rig Jacobian $\J$ changes over time.
Non-linear modifications of linear-blend skinning such as Wires \cite{SinghF98}, Dual-Quaternion
Skinning \cite{KavanCZO08}, Optimized Centers of Rotation \cite{LeH16}, Direct
Delta Mush \cite{LeL19} admit analytic --- albeit cumbersome --- derivatives
(see, e.g., \cite{GillesBFP11}).
Automatic differentiation 
provides a simpler more general solution at
approximately the same cost.
Most modern autodiff libraries provide APIs for computing Jacobians directly
giving a multi-variable function.

In a pinch, or if the rig function's implementation is not available, finite differencing
(cf.~\cite{hahn2012rig}) can gather approximate derivatives with simple forward
evaluation of the rig function. Looping over all rig parameters, we can
approximate the $j$th column of the 
the Jacobian with:
\begin{equation}
  \J_j \approx \frac{ \u^r( \rig_t + ε δ_j) -  \u^r( \rig_t - ε δ_j)}{2ε},
\end{equation}
where $δ_j ∈ \R^m$ is a vector of zeros except the $jth$ element is a one.

\begin{wrapfigure}[17]{r}{1.5in}%
  \includegraphics[width=\linewidth,trim={6mm 0mm 0mm 6mm}]{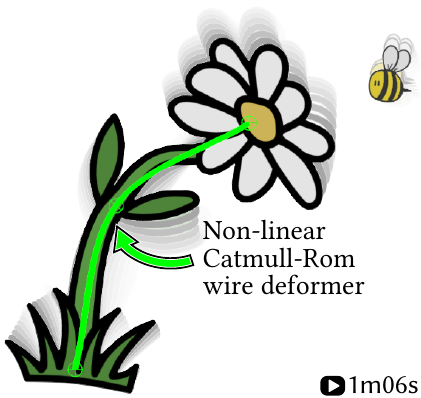}%
  \label{fig:daisy}
  \includegraphics[width=\linewidth,trim={6mm 0mm 0mm -2mm}]{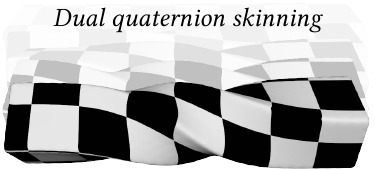}%
  \label{fig:dqs}%
\end{wrapfigure}
We reiterate that the rig Jacobian is computed \emph{once} per time-step, so
when optimizing for \emph{Non-Linear Elasticity} the rig Jacobian does not need
to be recomputed during each Newton iteration.
Hence, Jacobian computation is rarely the computational bottleneck.
The inset shows a non-linear wire deformer controlling the stem 
of a daisy,
while the pedals wiggle with secondary effects.

The inset shows dynamics added in the space orthogonal to the classic dual
quaternion skinning twisting bar setup.

\begin{figure}
\includegraphics[width=\linewidth]{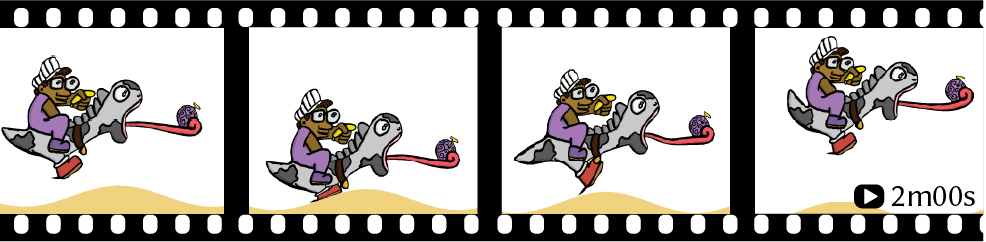}
\caption{\label{fig:mario} Rigidly directly 2D cartoon plumber riding a gray
dragon collides with the ground and our secondary effects act as elastic
responses. The dragon's tail and tongue wiggle with the induced momentum.}
\end{figure}

With various energy models and rig derivatives in hand, we have enough
information to bring rig animations to life with secondary effects (see
\reffig{amoeba}).
In \reffig{mario}, a 2D cartoon keyframed with affine motion collides with the
ground and impulses cause ripples through the shape without disrupting the input
path.

\subsection{Rig Momentum}
In the absence of external forces (e.g., winds or collisions), there will
be no secondary effect even if the rig moves.
This is a direct result of our separation of tasks: the artist controls
primary effects and physics \emph{in its own subspace} controls  separate
secondary effects.

Mathematically, we can see why momentum from the rig does not induce momentum
via the secondary displacements $\u^c$.
The object's momentum appears interacting with the rig displacements in the term
$\ud^\top \M \u^c$ in \refequ{Et}. 
Without loss of generality consider a linear rig, then the rig's momentum $\M
\ud^r$ can be written as $\M \J \dot{\rig}$,
so the rig's contribution to the momentum term is $\dot{\rig}^\top \J^\top \M
\u^c$.
However, the orthogonality constraint $\J^\top \M \u^c = 0$ causes this term to
vanishes and momentum does not pass between the rig and the simulation.

This strictly implements our contract with the artist.
On the other hand, in computer animation we often prefer overly floppy and
energetic characters.
Exaggeration and follow through principles of animation \cite{illusionoflife}
are easily accomplished with secondary effects coming from perceived internal
momentum.

Fortunately, we can clearly identify \emph{where} to inject momentum
from the rig into the dynamics.
By inserting a diagonal matrix $\D ∈ \R^{dn × dn}$ into the constraints
(replace $\J^\top \M \u^c$ with $\J^\top \M \D \u^c$ in \refequ{Jconst}), we can break the
annihilation in the $\ud^\top \M \u^c$ term and allow rig momentum to travel
through the shape into the secondary displacements.
This matrix acts as a map that transfers rig momentum to complementary dynamics 
momentum and any non-constant diagonal matrix, and most reasonable choices \emph{just work}. 

\begin{figure}
  \includegraphics[width=\linewidth]{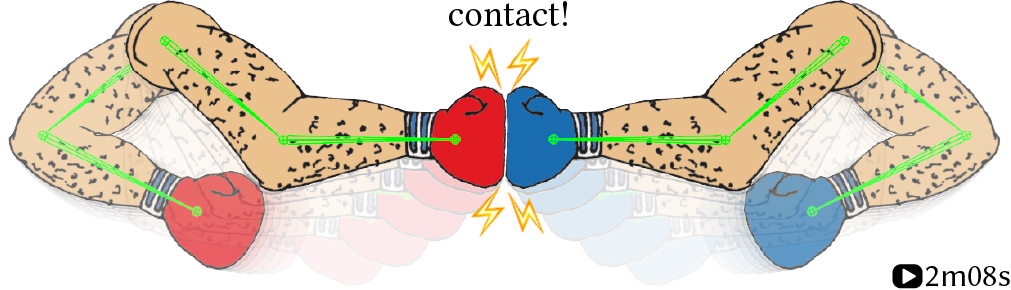}
  \caption{\label{fig:boxer-contact}
  The complementary dynamics of two rigged cartoons interact with each other
  through contact forces.
  }
\end{figure}

If and only if $\D = c \I$ where $c$ is a non-zero constant and $\I$ is the identity matrix, there will be no momentum interaction, as in all our examples 
with external forces (Figs.~\ref{fig:amoeba}, \ref{fig:mario}, \ref{fig:boxer-contact}, 
and inset in Sec. \ref{sec:results}). 
For all other examples, we construct a choice of $\D$ as a smooth transition via 
Poisson diffusion (i.e., $\mathbf{d} = 1$ on the surface and $∆ \mathbf{d} = 0$ on the interior).
For these, where $\D_{ii}=0$, we can expect rig momentum to ``leak'' into complementary 
dynamics momentum. 
This choice mimics cartoon-ish latency and follow-through \cite{illusionoflife}.
Because it is a dynamics simulation, this causes an elastic wave over the shape and quickly 
becomes high frequency ``wiggles'' \emph{everywhere}. 
So, $\D_{ii}=0$ can be seen roughly as the start locations of complementary dynamics waves.
Please note that, this is in stark contrast to painting softer/stiffer parts of
the mesh, since rig momentum will respect the underlying material of wherever their
induced wiggles travel. 
%
%

\section{Results}
\label{sec:results}
We have implemented our method in \textsc{Matlab} using \textsc{gptoolbox}
\cite{gptoolbox} 
and \textsc{Bartels} \cite{bartels} 
for geometry processing
routines and 
elastic energy derivatives, respectively. 
We did not optimize our code for performance and read in animations created in
\textsc{Maya} from file.
Our method has a similar overhead as past tracking methods
\cite{bergou2007tracks} which also augment the system matrix with linear
equality constraints.
Our prototype implementation uses \textsc{Matlab}'s sparse $LDL^T$ factorization
to solve the KKT system in \refalg{nm}.
We experimented with the null-space method but performance was slower as the
system matrix becomes dense even with a sparse null space basis (see, e.g.,
\cite{xu2016pose}).
We report the performance of our unoptimized \textsc{Matlab} prototype for
the examples in our paper in \reftab{timings}. 
Compared to unconstrained simulation (where the rig is entirely ignored, see Fig. 6), our current implementation shows a roughly 3$\sim$5X overhead
(\textsc{Matlab} switching from $chol$ to $ldl$ to handle our additional constraint) to achieve animation-driven simulation.
All timings are computed on a
MacBook Pro laptop with an Intel 2.3 GHz 8-Core i9 Processor and 16 GB RAM.

\begin{table}[t!]
\centering
\ra{1.2}
\setlength{\tabcolsep}{5.5pt}
\rowcolors{2}{white}{lightbluishgrey}
\begin{tabular}{l r r r r r r r r r r r r r r r r r r}
\rowcolor{white}
  Models & $n$ & $m$ & Avg. Frame \\
\midrule
  Roller Coaster (Inset of Sec.~\ref{sec:local-global})   & 39    & 6   & 0.003s  \\
  Non-angry bird (Fig.~\ref{fig:bird})                    & 639   & 18  & 0.38s  \\
  Carpet (Fig.~\ref{fig:carpet})                          & 239   & 18  & 0.12s  \\
  Worm (Fig.~\ref{fig:worm-propellor})                     & 438   & 6   & 0.05s  \\
  Daisy (Inset of Sec.~\ref{sec:non-linear-rigs})         & 702   & 6   & 0.42s  \\
  Monkey head (Fig.~\ref{fig:suzanne-tracks})             & 1244  & 12  & 2.86s   \\
  Plumber (Fig.~\ref{fig:mario})                          & 1502   & 6   & 0.07s  \\
  Bar (Inset of Sec.~\ref{sec:non-linear-rigs})           & 1519  & 16  & 4.61s   \\
  Walrus (Fig.~\ref{fig:walrus-cage})                     & 1559  & 18  & 0.14s  \\
  Hedgehog (Fig.~\ref{fig:hedgehog})                      & 1528  & 24  & 0.89s  \\
  Staypuft (Inset of Sec.~\ref{sec:results})              & 3258  & 72  & 359s    \\
  Cow (Bowl dropping) (Fig.~\ref{fig:spot-bob-drop})      & 3922  & 12  & 9.21s   \\
  Amoeba (Fig.~\ref{fig:amoeba})                          & 4918  & 6   & 0.06s  \\
  Fish (Fig.~\ref{fig:teaser})                            & 6142  & 24  & 14.8s   \\
  Elephant (Fig.~\ref{fig:cartoon-elephant})              & 8919  & 288 & 38.1s   \\
  TRex (Fig.~\ref{fig:trex})                              & 15623 & 600 & 49.2s   \\
\bottomrule
\end{tabular}
\caption{For each example we report the size of the simulation mesh $n$, the
  number of rig parameters $m$ and the average time spent computing
  complementary dynamics for an animation frame.}
\label{tab:timings}
  \vspace*{-0.9cm}
\end{table}

\begin{figure}
  \includegraphics[width=\linewidth]{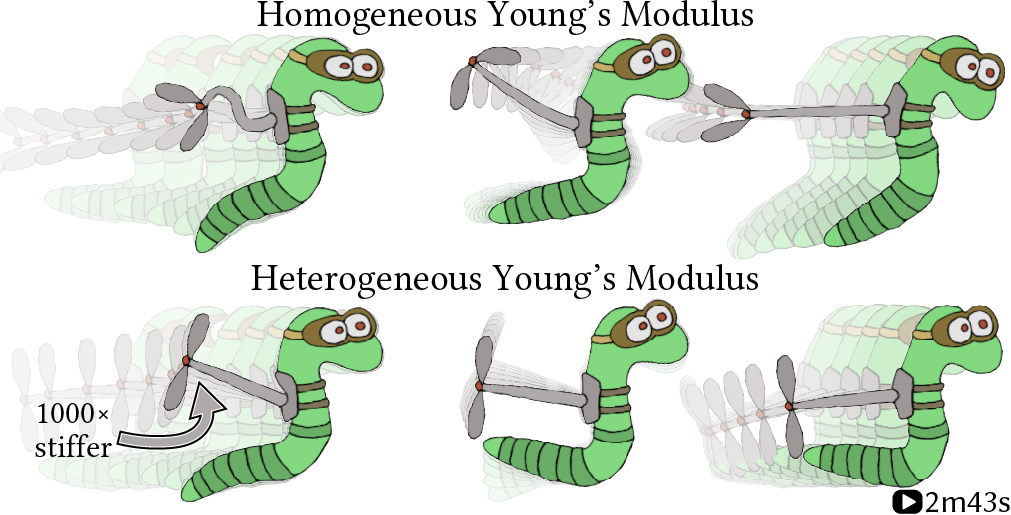}
  \caption{\label{fig:worm-propellor}
  Our method inherits advantages of whichever elasticity model it is
  plugged into (e.g., ARAP here \cite{ChaoPSS10}).
  Material parameters can be controlled just like a full-space simulation.
  }
  \end{figure}


In 2D, the input animation and simulation use the same triangle mesh.
In 3D,
for volumetric examples we tetrahedralize the domain to calculate physics
forces.
Depending on the example, we either use automatic skinning weights computed over
this tetmesh (e.g., \reffig{cartoon-elephant}) or for rigged models downloaded
off the internet (e.g., \reffig{trex}) we transfer the hand-painted skinning
weights to the tetmesh via bi-harmonic interpolation, conduct everything on the
tetmesh, and then display the embedded surface mesh.

Our complementary dynamics vastly simplify adding rich dynamic details to 
rig-based animations. 
Please see our supplemental video for animations of the following results as
secondary effects are difficult to capture in static images, but an essential
aspect to bringing moving objects to life \cite{illusionoflife}.

Our amoeba example~(\reffig{amoeba}) shows that even 2D animations, generated using only a 
single rigid handle can be significantly enhanced by our approach. 
Our method is not limited to simple forces,
\begin{wrapfigure}[8]{r}{1.5in}
  \includegraphics[width=\linewidth,trim={6mm 0mm 0mm 4mm}]{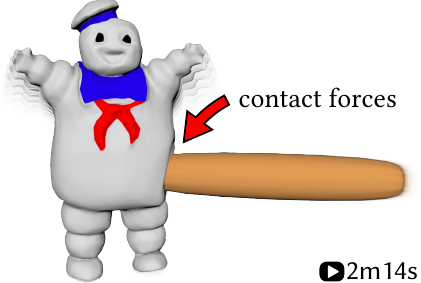}
  \label{fig:stay-puft}
\end{wrapfigure} 
%
%
but is also compatible with
off-the-shelf contact handling approaches (see \reffig{mario}).
For contact handling we rely on the force-based method of Heidelberger et
al~\shortcite{HeidelbergerTKMG04} (see inset).
Contact forces are computed and applied to mesh vertices
in the full space.
While our method (like many others) makes no formal guarantee about
finding a penetration-free solution, we quickly point out that our search
space is only \emph{slightly} smaller than the full space ($nd - m$).
This is a very different situation compared to subspace/rig-space physics (see
\refsec{rigspace})
whose search spaces are significantly reduced ($m$).
In \reffig{boxer-contact}, two rigged cartoons punch each other and the resolved
contacts cause secondary effects without disrupting the prescribed rig motion.
Our method enforces rig orthogonality with hard constraints. This might raise
fears of instability due to contradictory constraints or infeasibility,
for example when handling collisions. We have not observed such instabilities
or infeasibilities. We attribute this to the very large search space left to
the simulation, assuming the number of mesh vertices ($n$) is much larger than
the number of rig parameters ($m$).
The very fact that the rig alone cannot resolve contacts locally is an
indication that the constrained simulation \emph{can}. While it is
theoretically possible for an artist to create a large, highly constraining
rig that could impede collision resolution, the premise of complementary
dynamics is that the artist controls the high-level (and naturally
low-dimensional) motions and physical simulation synthesizes complex motions
independent of the rig complexity.

Since our method uses standard physics solvers to create secondary motion, any
contact handling scheme should be applicable.  Another advantage of our
physically-based approach is that heterogeneous object material parameters can
be tuned to produce desired results, such as stiffening the metal propeller on
the worm in \reffig{worm-propellor}.

\begin{figure}
  \includegraphics[width=\linewidth]{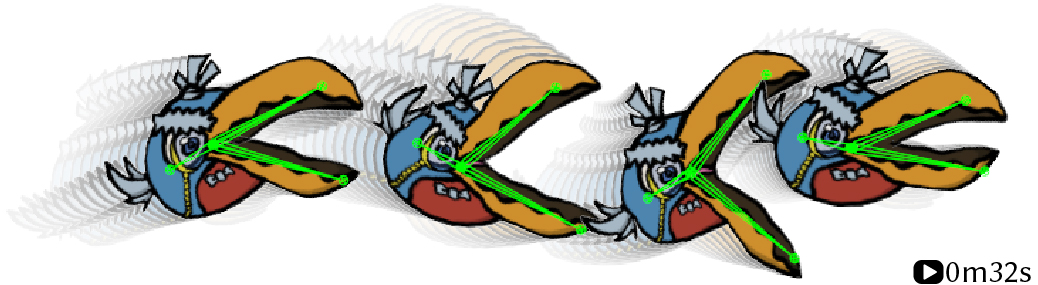}
  \caption{\label{fig:bird}
  An animator can focus on the primary rig motion (skeletal skinning, here) and
  our optimization adds secondary effects to create detailed motions.
  }
\end{figure}

One of the highlights of our method is its generality. 
It can bring lively secondary motion to most common rig types. 
As well as rigs involving multiple bones~(\reffig{bird}) and point handles~(\reffig{hedgehog}), 
our method works with cage-based deformers like this walrus~(\reffig{walrus-cage}) 
and even nonlinear rigs like this flower --- animated via Catmull-Rom wire
deformer~(daisy inset in \refsec{rigd}).
%

Naturally, these advantages extend to 3D.
Our teaser fish~(\reffig{teaser}) shows an example of a two bone rig.
Such a simple rig is able to instantly bring this fish to life with secondary
dynamics.
The animated T-Rex in \reffig{trex} demonstrates our ability to add
complementary dynamics to complex rigs found in the wild.

\begin{figure}
  \includegraphics[width=\linewidth]{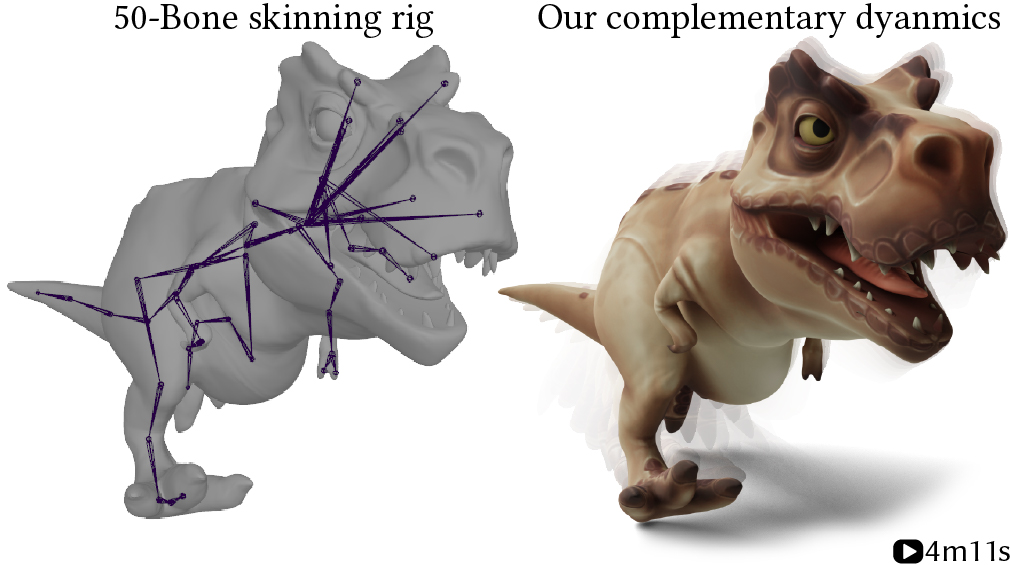}
  \caption{\label{fig:trex}
  This complex 50-bone linear blend skinning rig was downloaded from the internet
  \url{https://www.cgtrader.com/3d-models/character/other/toon-dinosaurs}. Despite
    the large rig space, our method still finds room for interesting
  secondary dynamics.
  }
  \end{figure}
\begin{figure}
  \includegraphics[width=\linewidth]{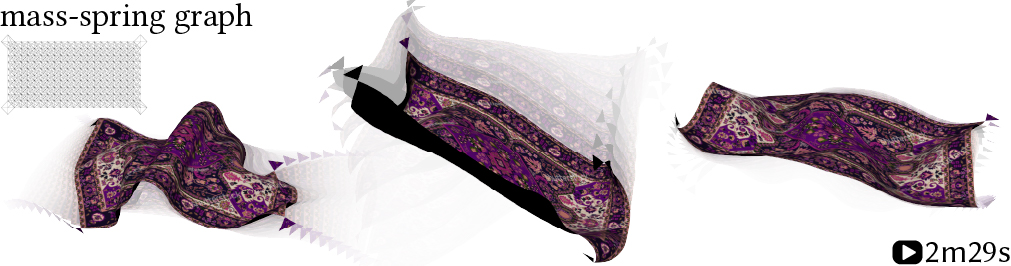}
  \caption{\label{fig:carpet}
  A carpet springs to life with complementary dynamics provided by a mass-spring
    cloth simulation 
    \cite{BW98}. Our constraints augment the local-global solver of
    \citet{liu2013fast}.
  }
\end{figure}

Our method exists
symbiotically with standard physics solvers.
This means, for instance, that we can use fast alternating solvers to create
complex cloth motions, as demonstrated by this magic ride~(\reffig{carpet}).
Our dancing elephant~(\reffig{cartoon-elephant}) illustrates how material
parameter tuning can be used to achieve the desired dynamic effect.

\begin{figure}
\includegraphics[width=\linewidth]{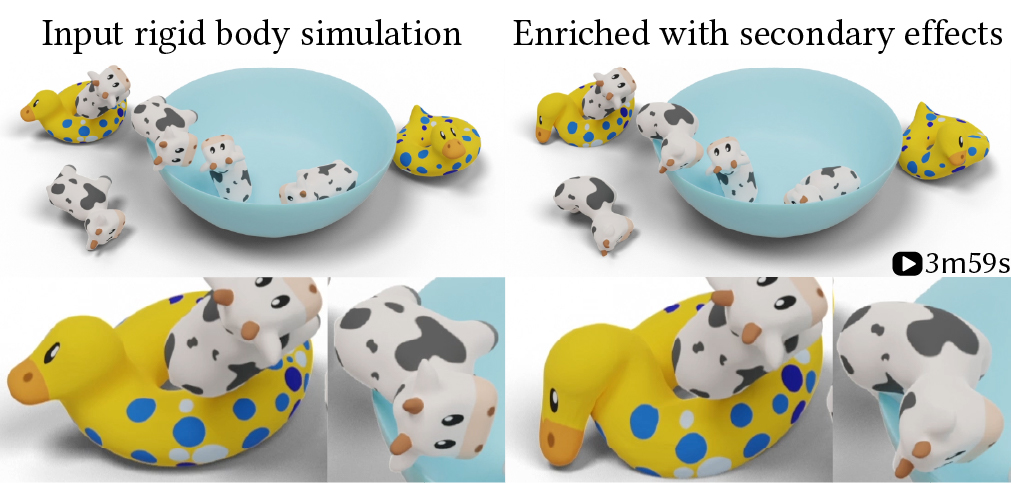}
\caption{\label{fig:spot-bob-drop} Our method does not rely on a rig. We can
  interpret each object of a rigid body simulation as being controlled by
  single-handle and then enrich the animation with secondary elastic effects
  while tracking the input closely. Compared to TRACKS, no segmentation is
  needed.
  }
\end{figure}

\begin{figure*}
  \includegraphics[width=\linewidth]{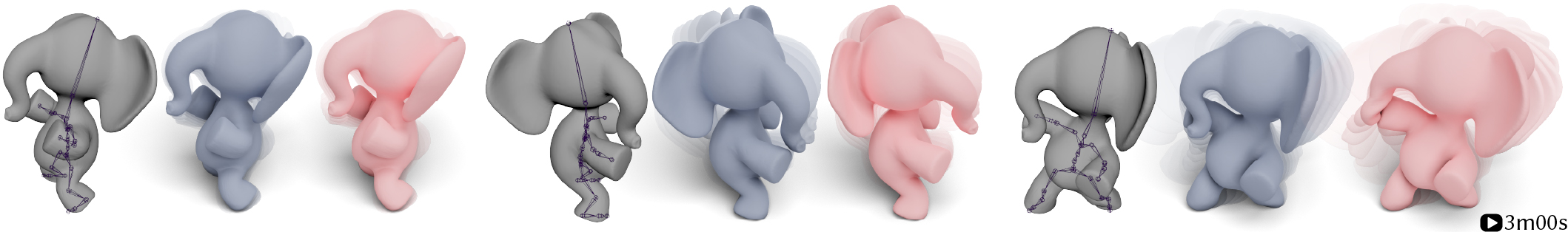}
  \caption{\label{fig:cartoon-elephant}
  A motion capture sequence controls the skeleton rigged inside of an elephant.
  The trunk and ears are not articulated by the skeleton so freely receive
  lively secondary effects.
  The material parameter (i.e., Young's modulus) of the blue elephant is set to
  be stiffer than the pink elephant.
  }
\end{figure*}

%
%
%

Rigid body simulations are easy for animators to set up and use to create
complex interactions with many objects.
However, the objects will looks stiff and lack elastic wiggling and jiggling.
Our method can treat each object of a rigid body simulation as a keyframed
animation and enrich the animation with exaggerated secondary effects (see
\reffig{spot-bob-drop}.
Unlike previous tracking methods (e.g., \cite{bergou2007tracks}), we do not
require a tuning a segmentation (see \reffig{suzanne-tracks}).

We treat a rigidly keyframed object as an \emph{Affine Body}. Technically this
increases the span of the rig-space (from $6$ to $12$ in $\R^3$) to include
scaling and shearing modes.
Our treatment is tantamount to assuming that the artist has intentionally held
those modes fixed to the identity.
We also experimented with true $6$-degree-of-freedom rigid keyframe rig function
(constructing $\J$ through an instantaneous exponential map), but the results
were similar with perhaps a bit more shearing in the dynamics.














\section{Conclusions and Future Work}
Our complementary dynamics bring a theoretically motivated, algebraic approach
to combining physics simulation with rigged animations created by an artist. 
By construction, our complementary dynamics cannot create motion inside of the ``rig-space'', and is thus
prevented from interfering with artistic intent. 
The algebraic nature of the method means it can be applied to a wide variety of rigs and 
make use of a wide variety of physics simulation algorithms -- as evidenced by the 
many results shown above.

Alas, complementary dynamics are not complimentary: they incur a computational
cost due to the additional constraints that must be applied during simulation. 
This paper focused on demonstrating generality with respect to the rig and
elastic model, rather than the well charted territory of performance
optimization. Nonetheless, it would be interesting in future work to see
complementary dynamics in a real-time setting and integrated with advanced
interfaces or live performance environment \cite{Willett:2017:SMF}, perhaps by
leveraging model reduction (cf., \cite{xu2016pose}).
Adding elastodynamics to an animation pipeline brings with it a host of constraints on input data.
The required mapping between well-behaved undeformed and deformed states means that 
our method (like all physics-based methods) is limited to well behaved geometry.
In general, making physics algorithms robust to pathological geometry (e.g
non-manifold or overlapping inputs~\cite{LiB18,XuB14}) is one of the most
important pursuits in the field of physics-based animation.

Currently our method does not support topological changes or artist prescribed
transitions between artworks~\cite{bai2016artist}, but extending our secondary
effects in this way is especially interesting for 2D animation.
Despite these limitations, our complementary dynamics is already a useful tool
for infusing rigged animations with life. Complementary dynamics turn physics
simulation into the artist's respectful partner, rather than an unruly party
crasher.

\subsection*{Acknowledgements}
This work is funded in part by NSERC Discovery (RGPIN–2017–05524, RGPIN-2017–05235,
RGPAS–2017–507938, RGPAS–2017–507909),
Connaught Fund (503114), CFI-JELF Fund, 
New Frontiers of Research Fund (NFRFE–201), the Ontario Early Research Award program,
the Canada Research Chairs Program,
the Fields Centre for Quantitative Analysis and Modelling and gifts by Adobe Systems,
Autodesk and MESH Inc.
We especially thank Paul Kry for hosting the 2018 Bellairs workshop on Computer Animation
and the attendees for inspiring initial project ideas.
We thank Otman Benchekroun, Rinat Abdrashitov and Josh Holinaty for proofreading;
John Hancock for the IT support;
anonymous reviewers for their helpful comments and suggestions.

\bibliographystyle{ACM-Reference-Format}
\bibliography{references}

\end{document}